\documentclass [10pt]{article}
\usepackage{graphicx}
\usepackage{amssymb} %(NOTA ! MI SERVIVA \lesssim E \gtrsim) (GP)
\usepackage{latexsym}
\usepackage{mathrsfs}
\usepackage{color}
\definecolor{red}{rgb}{1,0,0}

\makeatletter
\def\section{\@startsection {section}{1}{\z@}{-3.5ex plus -1ex minus
 -.2ex}{2.3ex plus .2ex}{\large\bf}}
\def\subsection{\@startsection{subsection}{2}{\z@}{-3.25ex plus -1ex
minus -.2ex}{1.5ex plus .2ex}{\normalsize\bf}}
\makeatother
\makeatletter

\@addtoreset{equation}{section}

\makeatother
\def\bea{\begin{eqnarray}} \def\eea{\end{eqnarray}}
\def\be{\begin{equation}} \def\ee{\end{equation}} \def\nn{\nonumber}
\def\a{& \hspace{-7pt}}  \def\Z{{\bf Z}}

\def\al{{\alpha^\prime}}

\def\thp{\theta^\prime}
\def\thpt{\tilde\theta^\prime}
\def\w{\omega}

\begin{document}

\thispagestyle{empty}

\begin{center}
\hfill SISSA-99/2005/EP \\
\hfill  HD-THEP-05-27 \\

\begin{center}

\vspace{1.7cm}

{\Large\bf A Note on T-duality in Heterotic String Theory }

\end{center}

\vspace{1.4cm}

{\bf Marco Serone$^{a}$ and Michele Trapletti$^{b}$}\\

\vspace{1.2cm}

${}^a\!\!$
{\em ISAS-SISSA and INFN, Via Beirut 2-4, I-34013 Trieste, Italy}

\vspace{.3cm}

${}^b\!\!$
{\em Institut f\"ur Theoretische Physik, Universit\"at Heidelberg,}\\
{\em Philosophenweg 16 und 19, D-69120 Heidelberg, Germany}

\end{center}

\vspace{0.8cm}

\centerline{\bf Abstract}
\vspace{2 mm}
\begin{quote}

We revisit the T-duality transformation rules in heterotic string theory,
pointing out that the chiral structure of the world-sheet leads to
a modification of the standard Buscher's transformation rules.
The simplest instance of such modifications arises for toroidal compactifications,
which are rederived by analyzing a bosonized version of the heterotic world-sheet Lagrangian.

Our study indicates that the usual heterotic toroidal T-duality rules naively extended
to the curved case cannot be correct, leading in particular to an
incorrect Bianchi identity for the field strength $H$ of the Kalb-Ramond field $B$.
We explicitly show this problem and provide a specific example of dual models where we
are able to get new T-duality rules which, contrary to the standard ones, lead to a
correct T-dual Bianchi identity for $H$ to all orders in $\al$.

\end{quote}

\vfill

\newpage

\section{Introduction}

Target-space duality symmetries, commonly denoted as T-duality symmetries, play
an important role in string theory (see \cite{Giveon:1994fu,Alvarez:1994dn} for a review on the subject).
They are part of the larger group of U-duality symmetries which relate in one way or an other
all M/string theories in a single underlying theory.
Being perturbative in the string loop expansion, T-duality is probably the best known
duality symmetry in string theory and the one which has better been established so far.
There are several ways to establish the transformation properties of massless fields,
the only relevant at low energies, under T-duality, both at the world-sheet level
\cite{Buscher,Rocek} and at the effective action
level (see e.g. \cite{Maharana:1992my}).

For the heterotic theory, however, things are more subtle due to the chiral structure
of the world-sheet. For instance, if one tries to naively implement the prescription
of \cite{Buscher} to get the transformation properties for the metric, the antisymmetric
Kalb-Ramond field and the gauge fields, one finds a set of transformations that do not
even match with the well-known $O(d,d+16)$ T-duality transformations of toroidal
heterotic backgrounds \cite{Hassan1,Maharana:1992my}
in the limit of flat space. In particular, the
dependence of the transformations from the Wilson lines
is not correctly reproduced. The reason for this failure is easy to understand.
Classical considerations such as those in \cite{Buscher} cannot detect
the presence of sigma-model anomalies \cite{Moore:1984dc,Hull:jv}
in the heterotic world-sheet Lagrangian.
This is also manifest from a space-time point of view, where it has been
shown that the low-energy effective action of heterotic toroidal compactifications
have an $O(d,d+16)$ symmetry, but only if one includes the gauge Chern-Simons (CS) corrections
in the definition of the field strength $H$ of the Kalb-Ramond field $B$ \cite{Maharana:1992my}.

Aim of this paper is to try to find the T-duality rules for heterotic theories
from a world-sheet point of view. Instead of dealing with anomalies and chiral fermions,
we find convenient to start from a bosonized version of the heterotic world-sheet
action, in which all subtleties related to anomalies are mapped to the subtleties
related to the dynamics of (anti)self-dual chiral bosons in two dimensions.
{}Following \cite{Alvarez:1994wj}, we analyze T-duality as a canonical transformation and
(re)derive the well-known toroidal T-duality rules in presence of Wilson lines.
We do it in two equivalent ways, which are related, in the fermionic version,
by a gauge transformation. The final transformations coincide only if one
assumes that the metric properly changes under a gauge transformation, as derived in \cite{Hull:1986xn}
by requiring world-sheet supersymmetry.

Our analysis clearly indicates that in curved spaces, even if no gauge
 background is switched on, the usual toroidal T-duality rules 
 cannot be naively applied (for example, as conjectured in \cite{Giveon:1991jj}). Indeed, they do not properly take into account the gravitational CS corrections entering in $H$. The latter is higher order in a derivative expansion
of the effective action, but is of the same order in $\al$ as the gauge CS correction
and is crucial to get a consistent Bianchi identity for $H$ in the T-dual background.
Extending our study to a generic curved space with or without
a gauge background appears to be difficult in general.
However, we have been able to provide an example of duality between two particular
T-dual curved backgrounds in which our analysis of the toroidal heterotic string can
straightforwardly be applied. The model is a generalization of the Melvin space of \cite{melvin},
already considered in this context in \cite{Serone:2003sv}. We explicitly show that, whereas
the ``naive'' T-dual rules would lead to the wrong Bianchi identity $dH=0$ for the T-dual background,
the corrected rules give exactly $dH = \al/4{\rm Tr} R(\omega)\wedge R(\omega)$ to all orders in $\al$,
where $\omega$ is the T-dual torsionfull spin connection, as computed directly by the duality,
and $H$ is the field strength of $B$, computed  as in \cite{Strominger:1986uh} from the complex structure
of the T-dual manifold. The generalized rules we have used in the curved case were actually
already conjectured in \cite{Bergshoeff:1995cg} from an analysis of $\al$ corrections in the low-energy action
of heterotic theories, for an arbitrary gauge and curved background.
Unfortunately, our world-sheet approach is not able to give a definite answer on the
validity of this assumption in general, which require further investigations.

This paper is organized as follows. In section 2 we briefly review the set-up in which
a T-duality transformation is seen as a canonical transformation.
In section 3 we extend this formalism to the heterotic string, allowing us to (re)derive
the T-duality rules for toroidal heterotic strings. In section 4 we apply our considerations
to a class of Melvin compactifications, showing how the Bianchi identity for $H$
is correctly reproduced in the T-dual background. Finally, in the appendix we report
the explicit transformations of the fields under a specific
$\Z_2$ subgroup of the T-duality symmetry.

\section{T-duality as a Canonical Transformation}

It has been proposed in \cite{Giveon:1988tt,Meissner:1991zj},
and more generally developed in \cite{Alvarez:1994wj},
that a T-duality transformation can be seen as a canonical transformation.
The essential idea is extremely simple and can conveniently be illustrated
in the easiest set-up of the closed bosonic string in flat space, compactified on a circle
of radius $R$.
We recall here how one gets the usual $\Z_2$ transformation $R\rightarrow \al /R$
in the canonical approach, following \cite{Alvarez:1994dn}.

Let us denote by $\theta$ the compact scalar (of period $2\pi$) associated to the
$S^1$ direction. The relevant world-sheet Lagrangian reads\footnote{Eq.(\ref{Lag-1})
is actually a Lagrangian density, related to the Lagrangian $L$ by $L=\oint\!d\sigma
\, {\cal L}$. Similarly, the Hamiltonian ${\cal H}$ introduced below is an Hamiltonian
density ${\cal H}$, related to the Hamiltonian $H$ by $H=\oint\!d\sigma
\, {\cal H}$. For simplicity, we will always refer to ${\cal L}$ and ${\cal H}$
as Lagrangian and Hamiltonian.}
\be
{\cal L} = \frac{R^2}{\al}\frac 12 (\dot\theta^2 -\theta^{\prime 2})\,,
\label{Lag-1}
\ee
where $\dot\theta=d\theta/d\tau$, $\thp=d\theta/d\sigma$.
Introducing the canonical momentum $P_\theta = \delta {\cal L}/\delta \dot\theta$, we get
the Hamiltonian
\be
{\cal H} = \frac 12 \frac{\al}{R^2} P_\theta^2 + \frac 12\frac{R^2}{\al}\,\theta^{\prime 2}
\,. \label{Ham1}
\ee
Consider now the following transformation of variables:
\be
\thpt  =  - P_\theta  \,, \,\,\,\,\, \\
\tilde P_{\tilde\theta}  =  - \thp \,.
\label{can-tran}
\ee
The transformation (\ref{can-tran}) is a canonical transformation, since the
structure of the Poisson brackets is preserved.
By expressing ${\cal H}$ in eq.(\ref{Ham1}) in terms of the dual variables, we get the
dual Hamiltonian:
\be
\tilde {\cal H} = \frac 12\frac{R^2}{\al}\,P_{\tilde\theta}^2+
\frac 12 \frac{\al}{R^2} \tilde\theta^{\prime 2}
\,. \label{Ham1t}
\ee
Define now $\dot{\tilde\theta}=\delta\tilde {\cal H}/\delta P_{\tilde\theta}$, so that
we get the dual Lagrangian $\tilde {\cal L}$:
\be
\tilde {\cal L} = \frac{\al}{R^2} \frac 12
(\dot{\tilde{\theta^2}} -\tilde\theta^{\prime\, 2})\,.
\label{Lag-dual-1}
\ee
Since the Lagrangians (\ref{Lag-1}) and (\ref{Lag-dual-1}) are related by a canonical
transformation, they represent equivalent descriptions of the same physical system.
Hence, the T-duality transformation $R\rightarrow \al /R$ is proved.

The generalization for a bosonic $\sigma$-model
\be
{\cal L} = \frac 12 G_{MN} (\dot x^M \dot x^N - x^{\prime \,M} x^{\prime \,N}) +
\frac 12 B_{MN} (\dot x^M x^{\prime \,N} -  \dot x^{N} x^{\prime \,M})
\label{Lag-2}
\ee
does not present any problem. T-duality can be implemented if there is a Killing vector $k^M$
under which the Lie derivative of the metric vanishes and $(i_k H)= k^M H_{MNP}$ is an exact
two-form, where $H=dB$ is the usual torsion tensor. In the so-called adapted coordinates,
$x^M=(\theta,x^i)$ and $k=\partial/\partial\theta$. Once written the $\sigma$-model (\ref{Lag-2})
in adapted coordinates, one defines the canonical momentum $P_\theta$ and the Hamiltonian
${\cal H}={\cal H}(P_\theta,\thp,x^i)$.\footnote{More precisely, since we are not introducing a
canonical momentum with respect to all coordinates, we are defining a Routhian and not
an Hamiltonian. Nevertheless, with an abuse of language, we will refer in the following to
the Hamiltonian.} The canonical transformation is still given by eq.(\ref{can-tran}),
through which one derives the dual Hamiltonian
$\tilde{\cal H} = \tilde{\cal H}(\tilde P_{\tilde \theta}, \thpt, x^i)$ and hence the dual Lagrangian
$\tilde {\cal L}={\cal L}(\tilde \theta, x^i)$. By comparing ${\cal L}$ with $\tilde {\cal L}$ (or equivalently
${\cal H}$ with $\tilde{\cal H}$)
one gets the well-known Buscher's relations \cite{Buscher}:
\bea
\tilde G_{00} & = & \frac 1G_{00}\,, \hspace{1.2cm} \tilde G_{0i} = \frac{B_{0i}}{G_{00}}\,,
\hspace{1.2cm} \tilde B_{0i} = \frac{G_{0i}}{G_{00}}\,, \nn \\
\tilde G_{ij} & = & G_{ij} - \frac{G_{0i}G_{0j}-B_{0i}B_{0j}}{G_{00}}\,,  \hspace{1.4cm}
\tilde B_{ij} = B_{ij} - \frac{G_{0i}B_{0j}-B_{0i}G_{0j}}{G_{00}}\,,
\label{buscher}
\eea
where $(0)$ labels the coordinate $\theta$.
Notice that eq.(\ref{can-tran}) gives the correct Buscher's T-duality rules only
with the normalization appearing in eq.(\ref{can-tran}).
In other words, if we denote $P_\theta$ and $\thp$ by $x_1$ and $x_2$, only if the
transformation $\tilde x_i= f_i(x_j)$ has unit Jacobian: $|\det (\partial_i f_j)|=1$.

Eqs.(\ref{buscher}) are valid for both compact and non-compact directions $x^i$.
In the former case, from an effective field theory point of view, the fields appearing in eq.(\ref{buscher})
are moduli scalar fields, and their T-duality transformations agree with those
found in \cite{Maharana:1992my} by an analysis of the effective field theory.
If the $x^i$ are non-compact, $G_{ij}$, $B_{ij}$, $G_{0i}$ and $B_{0i}$ are related to their
field theory counterparts $G_{ij}^{FT}$, $B_{ij}^{FT}$, $A_1\equiv G_{0i}^{FT}$ and $A_2\equiv B_{0i}^{FT}$ by known relations \cite{Maharana:1992my}, dictated by symmetries.
The gauge invariant quantities associated to these are $G_{ij}^{FT}$,
$F_G=d A_1$, $F_B=d A_2$ and $H=3dB^{FT}$.
In terms of these fields, the T-duality transformations act simply as
$\tilde G_{ij}^{FT}  =  G_{ij}^{FT}$, $\tilde A_1 = A_2$, $\tilde A_2 = A_1$,
$\tilde B_{ij}^{FT}  =  B_{ij}^{FT}$.
which reproduce, indeed, the transformations of \cite{Maharana:1992my} for the particular
$\Z_2$ transformation under consideration.
The generalization of the canonical formalism for superstrings has been worked out in
\cite{Hassan:1995je} and will not be reviewed here, since there are no substantial
modifications with respect to the bosonic case.

The canonical formalism reviewed above is probably the easiest approach to T-duality.
Its main drawback is in the transformation rule for the dilaton \cite{Buscher}
\be
\tilde \phi = \phi -\frac 14 \log \Big(\frac{G_{00}}{{\tilde G_{00}}}\Big)\,,
\label{dilaton}
\ee
which cannot be detected in a straightforward way.

\section{Constrained Systems and the Heterotic String}

The $\sigma$-model approach developed by Buscher to get the T-duality transformation rules
under $\Z_2$ isometries can also be extended to the heterotic string \cite{Alva,Bergshoeff:1994dg}.
In this case, one recovers exactly the transformations (\ref{buscher}) for $G_{MN}$ and $B_{MN}$, supplemented by
the transformations for the gauge fields $A_0$ and $A_i$:
\bea
\tilde A_0  =  -\frac{A_0}{G_{00}}\,, \,\,\,\,\,\, \\
\tilde A_i  =  -A_i + A_0 \frac{G_{0i}-B_{0i}}{G_{00}}\,.
\label{A-tran}
\eea
The Buscher's transformations (\ref{buscher}), as well as eqs.(\ref{A-tran}), are expected to receive
corrections in an $\al$ expansion (see e.g. \cite{Kal}). These depend on the particular background
and may also vanish, if the world-sheet has sufficient (super)symmetry.
In the heterotic case, instead, the transformations (\ref{buscher}) and (\ref{A-tran}) do not even reproduce the exact well-known T-duality rules for heterotic toroidal compactifications \cite{Hassan1,Maharana:1992my}.
One would argue that this is somehow related to the chiral structure of the heterotic
world-sheet action which necessarily requires a study of the chiral determinants
arising at one-loop level (in $\al$) due to the possible appearance of anomalies
(see e.g. \cite{Alva} for studies of world-sheet anomalies in this context).

Aim of this section is to compute at the world-sheet level
the above $\al$ corrections by considering the
bosonic formulation of the heterotic string in a canonical formalism.\footnote{Other works, such
as \cite{Maharana:1992my,Tseytlin:1990va,Schwarz:1993mg} have shown
how to implement T-duality in the world-sheet with (anti)self-dual scalars.
Their approach, however, was to find a suitable world-sheet action
which reproduced the known T-duality transformations rather than derive them from
$\sigma$-model considerations.}
As in section 2, we first consider a simple set-up, which
illustrates the main points we want to focus on: a compact scalar coupled with a $U(1)$ Wilson line.
The relevant degrees of freedom describing this system are two scalars: the compact scalar $\theta$
(of period $2\pi$) associated to a circle $S^1$ and a scalar $y$, associated to a $U(1)$ gauge symmetry.
The latter scalar is self dual, satisfying the constraint
$\dot y = y^\prime$, in real coordinates. Constrained systems can be treated in several
ways. In particular, \cite{Floreanini:1987as} (see also \cite{Faddeev:1988qp})
proposed a Lagrangian, linear in time derivatives,
for the description of a 2D self-dual field, which automatically incorporates the constraint.
Although this approach has been used with success in the past for the
description of the heterotic string (see e.g.
\cite{Tseytlin:1990va,Schwarz:1993mg}),
we find more convenient here to follow the canonical treatment of constrained systems
introduced by Dirac \cite{dirac}.

The relevant world-sheet Lagrangian reads
\be
{\cal L} = \frac 12 G (\dot\theta^2 -\theta^{\prime 2}) +
\frac 12 (\dot y^2 -y^{\prime 2}) + A (\dot y \thp - y^\prime \dot \theta)\,,
\label{Lag-Het-1}
\ee
where $G=R^2/\al$, $A$ represents a dimensionless $U(1)$ Wilson line along $S^1$
and the constraint $\Phi=\dot y - y^\prime=0$ is imposed, as usual, at the level of the
equations of motion only. Eq.(\ref{Lag-Het-1}) represents the bosonization of the
heterotic fermion Lagrangian
\be
{\cal L}_F = \frac 12 G (\dot\theta^2 -\theta^{\prime 2}) +
i\bar\lambda (\dot \lambda -\lambda^\prime) + A \bar \lambda \lambda (\dot \theta - \theta^\prime)\,.
\label{Lag-Het-F}
\ee
The constraint $\Phi=0$ is the main new feature
characterizing the canonical approach to the heterotic T-duality rules.
Its presence mixes the Poisson brackets of the various variables and correspondingly
we have to take canonical momenta both for $\theta$ and for $y$:
\be
P_\theta   = G \dot \theta - A y^\prime\, \,\,\,\,\, \\
P_y  =  \dot y + A \thp \,.
\ee
The Hamiltonian associated to the Lagrangian (\ref{Lag-Het-1}) reads
\be
{\cal H} = \frac 12 G^{-1} (P_\theta + A y^\prime)^2 + \frac 12 (P_y - A \thp )^2
+ \frac 12 y^{\prime 2} + \frac 12 G \theta^{\prime 2} \,.
\label{H0-het}
\ee
The constraint reads now
\be
\Phi\equiv P_y - A \thp - y^\prime =0\,.
\label{constraint-1}
\ee
Since this constraint is of second class, the only modification
that occurs is the replacement of the Poisson brackets by Dirac brackets, defined as
\be
\{A(\sigma),B(\sigma^\prime)\}_D = \{A(\sigma),B(\sigma^\prime)\}_P -
\int\! d\sigma^{\prime\prime} d\sigma^{\prime\prime\prime} \{A(\sigma),\Phi(\sigma^{\prime\prime})\}_P
C^{-1}(\sigma^{\prime\prime},\sigma^{\prime\prime\prime})
\{\Phi(\sigma^{\prime\prime\prime}),B(\sigma^\prime)\}_P,
\label{Dirac-bra}
\ee
where
\be
C(\sigma,\sigma^\prime)=\{\Phi(\sigma),\Phi(\sigma^\prime)\}_P\,.
\label{C-def}
\ee
The next step is to find a canonical transformation which gives the $\Z_2$  T-duality
transformations of $G$ and $A$. First of all, we look for a transformation which
leaves the Dirac brackets in the same form as the original ones.
It is not difficult to check that the following transformation satisfies this property:
\bea
\thpt & = & - P_\theta + A P_y \,, \hspace{1.9cm} \tilde P_{\tilde\theta}  =  - (1+A \tilde A) \thp - \tilde A y^\prime \,, \nn \\
\tilde y^\prime & = & - (1+A \tilde A) P_y + \tilde A P_\theta \,, \ \ \ \
\tilde P_{\tilde y}  =  - y^\prime - A \thp \,.
\label{dirac-1}
\eea
The transformation (\ref{dirac-1}) has unit Jacobian, as it should, and is parametrized
by $\tilde A$, the T-dual Wilson, which is so far unknown.
Thus, we will expect that eq.(\ref{dirac-1}) is not a canonical transformation
for any $\tilde A$. Rather, we fix $\tilde A$ by requiring that eq.(\ref{dirac-1})
is a canonical transformation.
We verify that by  showing that $H$ expressed
in the new variables has precisely the same functional form as in eq.(\ref{H0-het}).
By direct substitution, it is straightforward to verify that
\be
0=\Phi =  P_y - A \thp - y^\prime  =  \tilde P_{\tilde y} - \tilde A \thpt - \tilde y^\prime \equiv
\tilde \Phi =0 \,,
\label{phi-con}
\ee
where
\be
\tilde H(\tilde P_{\tilde y}, \tilde y^\prime, \tilde P_{\tilde\theta}, \thpt, A , \tilde A) = H(P_y,y^\prime,P_\theta,\thp,A)\,.
\label{h0def}
\ee
Eq.(\ref{phi-con}) shows that the form of the constraint is invariant under the transformation (\ref{dirac-1}).
Finally, one finds
\be
\tilde H_0(\tilde P_{\tilde y}, \tilde y^\prime, \tilde P_{\tilde \theta}, \thpt, A , \tilde A) =
H_0(\tilde P_{\tilde y},\tilde y^\prime,\tilde P_{\tilde\theta},\thpt,\tilde A)\,,
\label{h0Tdual}
\ee
provided that
\bea
\tilde G  =  \frac{G}{(G+A^2)^2}\,,\,\,\,\,\,\,  \\
\tilde A  =  - \frac{\al A}{(G+A^2)}\,,
\label{T-tran0}
\eea
or, in terms of the radii $R$ and $\tilde R$ ($\tilde G = \tilde R^2/\al$),
\bea
\tilde R  =  \frac{\al}{R(1+\frac{\al A^2}{R^2})}\,, \,\,\,\,\,\, \\
\tilde A  =  - \frac{A}{R^2(1+\frac{\al A^2}{R^2})}\,,
\label{T-tran1}
\eea
which more explicitly shows that the Wilson line corrections are sub-leading in an $\al$ expansion.
The T-duality transformations (\ref{T-tran1}) for the radius $R$ and for the Wilson line $A$
precisely agree with those expected in the heterotic case (see {\em i.e.} \cite{Hassan1},
where $A_{here}=A_{there}/2$). They can also be compared with the transformations given by
\cite{Maharana:1992my} by noting that eq.(\ref{T-tran0}) coincides
with the $\Z_2\subset O(2,1)$ transformation one gets from \cite{Maharana:1992my} for
\be
\Omega = \eta = \left( \matrix{ 0 & 1 & 0 \cr 1 & 0 & 0 \cr 0 & 0 & 1 \cr}\right)
\label{eta}
\ee
with the identification $A=a/\sqrt{2}$, in their notation.

By taking the inverse Legendre transformation, one can also check that
$\tilde {\cal L}=\tilde P_{\tilde y} \dot{\tilde{y}} + \tilde P_{\tilde \theta} \dot{\tilde{\theta}} - \tilde {\cal H}$,
where\footnote{In a constrained system, eqs.(\ref{Leg-Inv}) are not the Hamilton equations of motion, the latter being given by
$\dot{\tilde{\theta}} =\{\tilde \theta, \tilde {\cal H}\}_D$, $\dot{\tilde{y}} =\{\tilde y, \tilde {\cal H}\}_D$.}
\be
\dot{\tilde{\theta}} =  \frac{\partial \tilde {\cal H}}{\partial \tilde P_{\tilde \theta}} \,, \ \ \ \
\dot{\tilde{y}}  = \frac{\partial \tilde{\cal H}}{\partial \tilde P_{\tilde y}} \,,
\label{Leg-Inv}
\ee
has the same form as (\ref{Lag-Het-1}) in the T-dual variables, with $\dot{\tilde y}=\tilde{y^\prime}$.

The same physical system can alternatively be described in an other equivalent gauge.
If, before bosonizing the Lagrangian (\ref{Lag-Het-F}), we perform the non-single valued gauge transformation
$\lambda\rightarrow \lambda e^{i A \theta}$, we can set to zero the trilinear coupling in eq.(\ref{Lag-Het-F}).
This transformation is however anomalous \cite{Hull:jv} and it induces a shift in the metric
$G\rightarrow G+A^2$ \cite{Hull:1986xn}. In this gauge, the bosonized Lagrangian reads
\be
{\cal L} = \frac 12 (G+A^2) (\dot\theta^2 -\theta^{\prime 2}) +
\frac 12 (\dot y^2 -y^{\prime 2})\,,
\label{Lag-Het-2}
\ee
where now the constraint reads $\Phi=\dot y  - y^\prime+A (\dot \theta - \theta^\prime )=0$.
We can now proceed exactly as before, by replacing Poisson brackets with Dirac brackets.
Due to the absence of an explicit mixing between $\theta$ and $y$ in eq.(\ref{Lag-Het-2}),
the canonical transformations implementing the $\Z_2$ T-duality are trivial in this case:
\bea
\thpt & = & - P_\theta \,, \ \ \ \tilde P_{\tilde\theta}  =  -\thp\,, \nn \\
\tilde y^\prime & = &  P_y  \,, \ \ \ \ \
\tilde P_{\tilde y}  =   y^\prime\,.
\label{dirac-2}
\eea
The transformations (\ref{dirac-2}) lead to
\be
\tilde G +\tilde A^2 = \frac{1}{G+A^2}\,,
\label{GA-tran}
\ee
whereas the transformation of the Wilson line is obtained by
requiring that the dual constraint has the same form as the initial one. Indeed
\be
0 = \Phi = P_y - y^\prime + A \Big(\frac{P_\theta}{G+A^2}-\theta^\prime \Big)=
    \tilde \Phi =   \tilde y^\prime - \tilde P_{\tilde y} - \tilde A
\Big(\frac{\tilde P_{\tilde \theta}}{\tilde G+\tilde A^2}-\tilde \theta^\prime\Big) =0\,,
\ee
if $\tilde A$ is given by eq.(\ref{T-tran0}). Using the latter and
eq.(\ref{GA-tran}), the transformation
for $G$ appearing in eq.(\ref{T-tran0}) is reproduced.

The same procedure can be extended to a toroidal compactification on $T^n$ with
several $U(1)$ gauge fields. As before, the T-duality transformations can equivalently
be derived in one of the two gauges considered before.
In particular, the $\Z_2$ transformation $\Omega\subset O(n+16,n)$ under which one inverts the radii of all the torii,
$\Omega=\eta$, with $\eta$ the trivial generalization of the matrix (\ref{eta}), is obtained
exactly as before. One introduces canonical momenta for all the space-time bosons $\theta_M$ and all
the self-dual gauge scalars $y^a$ and consider the straightforward generalization of the
canonical transformation (\ref{dirac-1}). In this way, after some algebra, one ends
up with the correct $\Z_2$ T-duality transformation for a general toroidal compactification
of the heterotic string in presence of arbitrary Wilson lines and a constant background
for the Kalb-Ramond $B$--field.
In the gauge where all the couplings between the scalars $\theta^M$ and $y^a$ vanish
the Wilson lines enter in the redefinition of the metric
$G_{MN}\rightarrow G_{MN} + A_M^a A_N^a$ \cite{Hull:1986xn}
and of the constraint: $D_\tau y^a =\dot y^a + A^a_M \dot \theta^M = D_\sigma y^a = y^{a\prime} + A^a_M \theta^{M\prime}$.
The field $B_{MN}$, whose transformation is the key point of the Green-Schwarz anomaly cancellation
mechanism \cite{Green:1984sg,Hull:jv}, is left invariant since $B_{MN}\rightarrow B_{MN}+ A_{[M} A_{N]} = B_{MN}$.
In this way, the transformation for $G_{MN}$ and $B_{MN}$ arise from the canonical transformations, whereas
the ones for $A_M$ arise from the constraint.

For different $\Z_2$ isometries, such as the inversion of a single torus coordinate,
the T-duality transformations are not easily derived by the first method presented.
The subtlety is that one is expected to introduce canonical momenta only for the space-time scalar field $\theta$
associated to the isometry direction in which one wants to dualize the system, and not all of them. The constrained phase space is
given by the variables $\theta$, $P_\theta$, $y^a$ and $P_y^a$ and does not properly take into account the
possible presence of Wilson lines along the remaining directions. The scalars $x^i$, as far as the
symplectic space defined by the variables  $\theta$, $P_\theta$, $y^a$ and $P_y^a$ is concerned, are merely constants
and it turns out to be difficult to consistently treat the quadratic ``constant terms'' in $x^i$.
On the contrary, no subtleties arise in the gauge in which $D_\tau y^a  = D_\sigma y^a$.
The transformation for the metric and the $B$--field are obtained by comparing the original and
dual Hamiltonians or Lagrangians, whereas the transformation for the Wilson
lines $A_0^a$ and $A_i^a$ are derived by comparing
the two constraints. We report in an appendix the explicit form of the T-duality transformations
for this $\Z_2$ isometry.

The transformations (\ref{T-dual-expl}), like eqs.(\ref{buscher}), are valid for both compact and non-compact
directions $x^i$.
In the former case, all fields appearing in eq.(\ref{T-dual-expl})
are moduli scalar fields, the constant $A_0^a$ and $A_i^a$ are true Wilson lines
and their T-duality transformations agree with those
found in ref.\cite{Maharana:1992my} by an analysis of the effective field theory.
If the $x^i$ are non-compact, the backgrounds $A_i^a$ are constant which can be gauged away.
The presence of these fields, however, modify the relation between
$G_{ij}$, $B_{ij}$, $G_{0i}$, $B_{0i}$  and their
field theory counterparts $G_{ij}^{FT}$, $B_{ij}^{FT}$, $A_1$ and $A_2$, as
shown in \cite{Maharana:1992my}. {} From a field theory point of view, the non-trivial structure of
the $\Z_2$ T-duality transformation encoded in the gauge connections is totally reabsorbed in the relation between
the world-sheet fields and their field-theory partners. Indeed, the only non-trivial T-duality
transformations for the field theory fields are $\tilde A_1=A_2$, $\tilde A_2=A_1$,
since also the gauge fields are left invariant: $\tilde A_i^{FT} = A_i^{FT}$.

It should be clear from the above world-sheet derivation, that the
T-duality rules (\ref{T-dual-expl}) are strictly valid only for {\it constant connections}
and {\it flat spaces}. Indeed, whereas in Type II theories the transformations (\ref{buscher})
are subject only to $\al$ corrections related to the renormalization of the world-sheet action,
the chiral nature of the heterotic world-sheet represents an other source of corrections.

\section{Curved Background: a Concrete Example}

We focus on a special class of curved backgrounds where the spin-connection is constant,
in complete analogy to a Wilson line,
and set to zero the gauge and $B$--field background. For simplicity, we switch in
this section our conventions for left and right moving fields
with respect to the last section, so that
the ``space-time" world sheet fermions here have the same chirality
as the ``gauge'' fermions in the last section.
The spin connection components $\omega_0^{ab}$ and
$\omega_i^{ab}$,  where the subscript index is curved and the upper ones are flat,
are taken to be  constant and already in a skew-symmetric form.
We also require the existence of a killing vector
$k=\partial/\partial \theta$.
The analysis of such a background  is similar to that performed for the toroidal model in the last section;
by a non-single valued Lorentz
transformation, we can set to zero $\omega$ producing a
shift in the metric term $G_{MN}\rightarrow G_{MN}+\omega_M^{ab} \omega_N^{ab}$ \cite{Hull:1986xn}.
We can introduce the canonical transformation (\ref{dirac-2}) and read
out the form of the dual spin connection (in this gauge) by the transformation of the
constraint.
Given this, we can recover the dual metric by extrapolation from
(\ref{GA-tran}), and completely describe the dual background via the
transformations (\ref{T-dual-expl}).
Differently from the toroidal case, the T-dual spin connection is not
constant and the space-time has a non-trivial curvature and torsion.
The new (fermionic) action is given by eq.(7) of \cite{Hull:1986xn},
in presence of a non-trivial dual spin connection $\tilde\omega\equiv\omega^{(-)}$,
as given by (\ref{T-dual-expl}), and $\omega^{(+)}=0$, in the notation of \cite{Hull:1986xn}.

Freely acting orbifolds furnish a concrete example of backgrounds
with this feature.
Consider in particular spaces of the form $S^1\times (\mathbb C^n\times S^1)/\Z_N$.\footnote{We analyze $\mathbb C^n$ and not $T^n$ because the T-duality transformation we are considering does not admit a consistent realization on $T^n$.}

We parametrize the space $\mathbb C^n\times S^1\times S^1$ with
$n$ complex variables $w_i\in\mathbb C$,  $i=1,\dots,n$, and two real
variables $\theta\in[0,2\pi)$ $x_1\in[0,2\pi)$.  The parameters are
dimensionless and we introduce a dimensionfull metric
\bea
ds^2=R^2 d\theta^2+R^2 dx_1^2+\al\,dw_i d\bar w_i.
\eea
The orbifold group is $\Z_N=\{I,g,g^2,\dots g^{N-1}\}$, and the action
of $g$ is
\bea
g:\left\{
\begin{array}{l}
\theta\rightarrow \theta+2\pi/N\\
x_1\rightarrow x_1\\
w_i\rightarrow e^{2 \pi i \beta_i/N} w_i
\end{array}
\right.,
\eea
with $\beta_i\in\mathbb N$. The $S^1$ parametrized by $x_1$ is just a
spectator direction, introduced in order to have the possibility
of defining a complex structure.
Since the orbifold action is free, $g$ has no fixed points,
$S^1\times (\mathbb C^n\times S^1)/\Z_N$ is a smooth manifold, a
fibration of the complex planes over $S^1$.
We can define new parameters
$z_i=w_i\, \exp(-i\beta_i\,\theta)$,
in which the orbifold action is trivial.
It is useful to introduce real coordinates by defining
$z_i=x_{2i}+i x_{2i+1}$. The metric now reads
\be
ds^2=R^2 dx_1^2+\al\rho_0\,d\theta^2+\al\sum_{i=2}^{2n+1} dx_i^2+
2\al\,d\theta \sum_{j=1}^n\beta_j (x_{2j} dx_{2j+1}-x_{2j+1} dx_{2j}),
\label{metric}
\ee
where
\bea
\label{defro}
\rho_\chi=\frac{R^2}{\al}+\sum_i\beta_i(x_{2i}^2+x_{2i+1}^2)+\chi,
\eea
the constant
$\chi$ having been introduced for later convenience. The non-zero
terms of the spin connection $\w_{\mu}^{ab}$ are $\w_0^{23}=\beta_1$,
$\w_0^{45}=\beta_2$, $\w_0^{67}=\beta_3$, etc., i.e.
$\w_0^{(2j)(2j+1)}=\beta_j$.
The manifold is locally flat but with a non-trivial constant spin connection
and a non-trivial $\Z_N$ holonomy group.
It is a generalization
of the backgrounds introduced in \cite{melvin}.

Consider now the heterotic string on such spaces.
The relevant world-sheet fields are $2n+2$ bosons $x^i$
and their left-moving fermion partners $\psi^i$.
The fermion Lagrangian reads
\bea
\label{newspinor}
{\cal L}=i\psi^a\partial_+\psi^a -
\partial_+ \theta\,\w_0^{ab}\,\psi^a\psi^b\,,
\eea
where $\partial_+ \psi^a=\dot \psi^a-\psi^{\prime a}$ and $a=2\dots 2n$
(the $\psi^0$ and $\psi^1$ fermions are decoupled).
Eq.(\ref{newspinor}) can be bosonized, since the spin connection
$\w_0^{(2j)(2j+1)}=\beta_j$ couple $\partial_+\theta$
to the fermions $\psi^{2j}$ and $\psi^{2j+1}$ only.
This dynamics is completely reproduced by $n$ constrained
bosons $y^j$ coupled with $\theta$ via $\beta_j$ itself, in the Lagrangian term
$\beta_j (\dot \theta {y^i}^\prime- {\theta}^\prime {\dot y^i})$.

There are several embeddings of the orbifold
action in the gauge degrees of freedom which are consistent with modular invariance.
Their net effect is encoded in a non-vanishing gauge connection, of the form $A_0^{(2j)(2j+1)}=\alpha_j$,
with $\alpha_j\in\mathbb N$.
Unfortunately, we are not able to describe the T-duality
in presence of both non-zero spin and gauge connections, so the favorite choice
will be $\alpha_i=0$. This is consistent with modular invariance only if
$\sum_i \beta_i^2=2N \,{\rm mod}\, 2N$; for instance, with $n=3$
and $N=3$, $\beta_i=(1,1,-2)$ or, with $N=7$, $\beta_i=(1,2,-3)$.
From here on we especially refer to these choices, which also lead to space-time
supersymmetry, but we keep a generic notation, leaving $N$, $n$ and $\beta_i$ as unspecified parameters.

We can consider a $\Z_2$ T-duality transformation on this system, along the isometry direction
$\theta$. First of all, it is useful to write down the dual backgrounds
obtained by using the standard T-duality rules (\ref{buscher}). We get
the following dual metric, $\tilde B$--field and dilaton backgrounds:
\bea
d\tilde s & = & R^2 dx_1^2+\al\rho^{-1}_0 d\theta^2+\al \sum_{i=2}^{2n+1} dx_i^2-
\al\rho^{-1}_0
\left\{\sum_{j=1}^n\beta_j (x_{2j} dx_{2j+1}-x_{2j+1} dx_{2j})\right\}^2, \nn \\
\tilde B_{0(2j)}& = & \beta_j \,x_{2j+1}\,\,\rho^{-1}_0, \hspace{1cm}
\tilde B_{0(2j+1)}=-\beta_j \,x_{2j}\,\,\rho^{-1}_0, \hspace{1cm}
\tilde \phi  =  -\frac{1}{2} {\rm Log}\left(\rho_0\right),
\label{Tdualcu}
\eea
while the gauge fields are all vanishing.
The manifold admits a globally-defined complex structure $\tilde J$, whose
explicit form can easily be computed using the T-dual map
for complex structures as given in \cite{Hassan:1994mq}.
Following \cite{Strominger:1986uh}, we can compute $\tilde H$
from $\tilde J$ and in this way one can check that $\tilde H=d\tilde B$, with
the $\tilde B$--field as in eqs.(\ref{Tdualcu}).
Moreover, it is possible to check that the background is an exact
solution of the equations of motion of 10d SUGRA, as described  in
\cite{Strominger:1986uh}.
Unfortunately, there is a failure \cite{Serone:2003sv}: since $\tilde H=d\tilde B$, it follows that
$d\tilde H=0$, and the Bianchi Identity $d\tilde H=\frac{\al}{4\pi}({\rm Tr}
\tilde R\wedge \tilde R-{\rm Tr} \tilde F\wedge \tilde F)$, is not fulfilled since $\tilde F=0$
but ${\rm Tr} (\tilde R(\tilde \omega)\wedge \tilde R(\tilde \omega))\neq 0$. The latter quantity does not vanish
independently of the choice of connection, namely for
$\tilde\omega=\tilde\omega_0+ c \tilde H$, where $\tilde \omega_0$ is the torsion-free
spin connection and $c$ is an arbitrary constant.

We can now pass to the study of the T-duality with corrections.
Using eqs.(\ref{T-dual-expl}), we get
\bea
d\tilde s& =& R^2 dx_1^2+\al\frac{\rho_0}{\rho_\chi^2} d\theta^2-
2\al\frac{\chi}{\rho_\chi^2} d\theta
\left\{\sum_{j=1}^n\beta_j (x_{2j} dx_{2j+1}-x_{2j+1} dx_{2j})\right\}+ \nn \\
\a\a
\al \sum_{i=2}^{2n+1} dx_i^2-
\al\frac{\rho_\chi-\chi}{\rho_\chi^2}
\left\{\sum_{j=1}^n\beta_j (x_{2j} dx_{2j+1}-x_{2j+1} dx_{2j})\right\}^2,  \nn \\
\tilde B_{0(2j)}& = & \beta_j \,x_{2j+1}\,\,\rho_\chi^{-1}, \ \ \ \
\tilde B_{0(2j+1)}=-\beta_j \,x_{2j}\,\,\rho_\chi^{-1}, \ \ \ \
\tilde \phi=-\frac{1}{2} {\rm Log}\left(\rho_\chi\right),
\label{ExBack}
\eea
where $\rho_\chi$ has been defined in (\ref{defro}) and
$\chi=\sum_{i=1}^n \frac{1}{2} \beta_i^2$
is the term $A_0^2$ appearing in eqs.(\ref{T-dual-expl}).
Notice that in the dual background (\ref{ExBack}), which at leading order coincides with (\ref{Tdualcu}),
the metric components $G_{0i}$ are not vanishing as in eq.(\ref{Tdualcu}).
The metric in (\ref{ExBack}) admits a new complex structure $\tilde J$ from
which we can deduce the torsion $\tilde H$.
Now there is a mismatch between $\tilde H$
and $d\tilde B$, precisely keeping into account the presence
of the gravitational terms in the Bianchi Identity.
More in detail, having the dual spin connection $\tilde \omega^{(-)}$ directly
from eq.(\ref{T-dual-expl}), we can check that the Bianchi Identity
\be
d\tilde H = \frac{\al}{4} {\rm Tr}\, \tilde R(\tilde \omega^{(-)})\wedge \tilde R(\tilde \omega^{(-)})\,,
\ee
is satisfied to {\it all} orders in $\al$.
Moreover, the background (\ref{ExBack})  satisfies exactly the 10d SUGRA equations of motion.

We would like to point out that the  corrections to T-duality rules
we are proposing do not simply boil down to the Buscher ones on a shifted metric,
since the direct transformations of the spin connection $\omega^{(-)}$ appearing in eq.(\ref{T-dual-expl}) are
also needed. It should also be clear that our formalism do not simply work for
a general pair of dual curved spaces, where both spaces have a non-vanishing
curvature, indicating that it is difficult to imagine that the T-duality rules
(\ref{T-dual-expl}) are somehow valid also in such cases, as conjectured in \cite{Bergshoeff:1995cg}.
An exception is represented by compactifications with standard embedding. In such a case,
being the underlying world-sheet effectively vector-like, we expect that all the
modifications to Buscher's rules we have been considering will eventually vanish.
It would be interesting to check and possibly generalize from the low energy effective action point of view
the modified T-duality rules we propose, developing more on \cite{Bergshoeff:1995cg}.

\section*{Acknowledgements}

We thank L. \'Alvarez-Gaum\'e and K. Narain for several interesting discussions and
comments at the early stages of this work. We also thank G. Bonelli for useful discussions.
M.T. thanks the hospitality of SISSA where part of this project was developed.
Work partially supported by the European Community's Human
Potential Programme under contract MRTN-CT-2004-005104 and by
the Italian MIUR under contract PRIN-2003023852;
work of M.T. supported by the DFG (German Science Foundation)
under contract HE 3236/3-1.

\appendix

\section{Heterotic T-duality along $S^1$}

We report in the following the explicit form of the $\Z_2$ T-duality
transformations along a circle for metric, $B$--field and gauge fields
for the toroidal compactification of the heterotic string.
For completeness, we also report here the known transformation rule of the dilaton, although
the canonical approach does not allow to fix it:
\bea
\tilde G_{00} & = & \frac{G_{00}}{(G_{00}+A_0^2)^2}\,, \nn \\
\tilde G_{0i} & = &
\frac{G_{00} B_{0i}+ (A_0)^2 G_{0i}- A_0\cdot A_i G_{00}}{(G_{00}+A_0^2)^2} \,, \nn \\
\tilde G_{ij} & = & G_{ij}-\frac{G_{0i}G_{0j}-B_{0i}B_{0j}}
{G_{00}+ A_0^2}- \nn \\
& & \frac{1}{(G_{00}+A_0^2)^2}
\left\{G_{00}\left[B_{0j}A_0 \cdot  A_i +B_{0i} A_0 \cdot  A_j
-(A_0 \cdot A_i) (A_0 \cdot A_j)\right]\right. \nn \\ &&
\left.
+A_0^2 \left[ (G_{0i}-B_{0i})(G_{0j}-B_{0j})+
(G_{0i}A_0 \cdot  A_j+G_{0j} A_0 \cdot  A_i)
\right]
\right\} \,,\nn \\
\tilde A_0^a & = &  - \frac{A_0^a}{G_{00}+A_0^2}\,, \nn \\
\tilde A_i^a& = & - A_i^a + A_0^a\frac{G_{0i}-B_{0i}+A_0 \cdot A_i}{G_{00}+A_0^2}\,, \nn \\
\tilde B_{0i} & = & \frac{G_{0i}+ A_0 \cdot A_i}{G_{00}+A_0^2} \,, \nn \\
\tilde B_{ij} & = & B_{ij}-
\frac{(G_{0i}+A_0\cdot A_i) B_{0j}-
(G_{0j}+A_0\cdot A_j) B_{0i}}{G_{00}+A_0^2}\,, \nn \\
\tilde \Phi & = &\Phi + \frac 14\log\bigg[\frac{{\rm Det} \tilde G}{{\rm Det } G}\bigg] \,.
\label{T-dual-expl}
\eea
In eq.(\ref{T-dual-expl}), $A_0^2 = A_0^a A_0^a$, $A_i\cdot A_j = A_i^a A_j^a$.
The $\al$-dependence can be made explicit by replacing the dimensionless metric and
$B$--field with the dimensionfull ones $G\rightarrow G/\al$, $B\rightarrow B/\al$.
In this way it is clear that the terms $A_0^2$ and $A_0\cdot A_i$ in eq.(\ref{T-dual-expl})
represent $\al$ corrections to the standard T-duality rules.
The gravitational case is taken into account by simply replacing $A$ with the
spin connection $\omega^{(-)}$.

\end{document}